\documentclass[final,5p,times,twocolumn]{elsarticle}

\usepackage{amssymb} % Allows line breaking in URLs
\usepackage[hyphens]{url}  % Better hyphenation in URLs
\usepackage{breakurl}
\usepackage{hyperref}
\usepackage{etoolbox} 
\usepackage{balance}
\usepackage{amsmath}
\usepackage{array}
\newcommand{\orcidlink}[1]{\textsuperscript{\href{https://orcid.org/#1}{\includegraphics[scale=0.2]{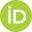}}}}

\begin{document}

\begin{frontmatter}



\title{Pushing the Boundaries of Immersion and Storytelling: \\A Technical Review of Unreal Engine}

\author[label2]{Oleksandra Sobchyshak\,\orcidlink{0009-0000-7006-7487}}
\author[label2]{Santiago Berrezueta-Guzman\,\orcidlink{0000-0001-5559-2056}}
\author[label2]{Stefan Wagner\,\orcidlink{0000-0002-5256-8429}}
\affiliation[label2]{organization={Technical University of Munich},
           city={Heilbronn},
            country={Germany}}

\begin{abstract}
Unreal Engine is a platform that has influenced immersive storytelling and virtual reality (VR) through its advanced features and diverse applications. This paper provides an in-depth technical review of Unreal Engine. It analyzes its key innovations in creating hyper-realistic environments and emotionally engaging narratives, with significant applications in gaming, virtual production, education, cultural preservation, and healthcare.
The findings of this article highlight Unreal Engine’s transformative impact across industries, demonstrating its ability to merge storytelling with cutting-edge technologies. Case studies illustrate how Unreal Engine facilitates seamless visuals, audio, and interactivity integration to create compelling experiences. Additionally, this study identifies Unreal Engine’s versatility in applications ranging from procedural content generation and AI-driven workflows to smart-city simulations and VR-based rehabilitation programs.

While Unreal Engine sets new benchmarks for visual fidelity and interactivity, this paper underscores critical challenges, including its high hardware demands, limited accessibility, and ethical concerns related to over-immersion and data privacy. Addressing these challenges through cloud-based rendering, inclusive design, and ethical practices is essential for broader adoption and sustainability.
This review concludes that Unreal Engine is suitable for innovation and interdisciplinary collaboration. Its ability to empower creators, redefine workflows, and push the boundaries of immersive storytelling positions Unreal Engine as pivotal in shaping the future of virtual reality and interactive media.

\end{abstract}

\begin{keyword}

Unreal Engine \sep Virtual Reality\sep Immersive Storytelling\sep Nanite\sep Lumen\sep Virtual Production\sep Cultural Preservation\sep Dynamic Lighting\sep Game Development\sep Accessibility\sep Ethical Design

\end{keyword}

\end{frontmatter}

\section{Introduction}\label{I}

Virtual Reality (VR) revolutionizes storytelling by enabling immersive, interactive narratives that engage users on a personal, multi-sensory level. It enhances emotional connections, allows exploration and influence of the story, and expands applications in education, entertainment, and therapy \cite{ryan2015narrative, chen2025task}. 
However, immersive storytelling presents significant technical challenges that must be addressed for widespread adoption. One primary challenge is the real-time rendering of complex environments while maintaining performance efficiency, particularly for VR applications that require high frame rates to prevent motion sickness \cite{mack2019unreal}. Another challenge is ensuring dynamic interactivity without sacrificing visual fidelity, as traditional pre-rendered assets are often too rigid for fluid storytelling \cite{bowman2007virtual}. Additionally, spatial audio precision and realistic character animation are critical for immersion but require sophisticated algorithms to achieve real-time adaptability \cite{broderick2018importance}.

The evolution of game engines has been instrumental in addressing these challenges through advanced rendering innovations. Unreal Engine, developed by Epic Games, has evolved from a high-performance rendering tool primarily used in the gaming industry to a versatile platform driving innovation across multiple fields, including virtual production, cultural preservation, and healthcare. Since its release in 1998, Unreal Engine has undergone significant advancements, with Unreal Engine version 5 introducing technologies such as Nanite for high-detail asset rendering, Lumen for real-time global illumination, and MetaHuman Creator for lifelike character animation \cite{silicoptimizing, el2022unreal}.

Beyond its technical excellence, Unreal Engine's importance lies in providing tools for developers, storytellers, and educators to create captivating experiences by simplifying workflow and making advanced tools accessible without requiring extensive technical knowledge or resources. With the adoption of Unreal Engine comes challenges such as high demands on hardware and ethical considerations in content creation \cite{dooley2023creating}. 

This paper provides a comprehensive technical review of Unreal Engine, systematically examining its core principles, key technologies, applications, and limitations. By analyzing its real-time rendering innovations, interactive storytelling techniques, and computational efficiency, we aim to highlight Unreal Engine’s strengths while addressing accessibility challenges and ethical considerations.

The structure of this paper is as follows: Section \ref{CF} highlights core technologies like Nanite, Lumen, and MetaHuman Creator. Section \ref{S} explores Unreal Engine's role in immersive storytelling, while Section \ref{A} examines its applications across gaming, virtual production, and cultural preservation. Sections \ref{T} and \ref{COM} detail narrative techniques and compare Unreal Engine with other engines. Challenges, ethical considerations, and future directions are discussed in Sections \ref{CH} and \ref{F}, with conclusions summarized in Section \ref{C}.

\section{Core Features of Unreal Engine}\label{CF}

Unreal Engine represents the revolutionary leap in creating and rendering digital content. This section classifies its core technological components and explains their underlying principles and algorithms, highlighting their impact on immersive storytelling and computational efficiency \cite{sanders2016introduction}.

\subsection{Nanite Virtualized Geometry}

Nanite introduces a data-driven rendering pipeline that dynamically optimizes asset resolution based on viewing distance and camera perspective. Unlike traditional Level of Detail (LOD)-based rendering, where multiple versions of an asset must be created manually, Nanite automatically adjusts polygon density without losing detail. This is achieved through Hierarchical Level-of-Detail (HLOD) algorithms, which cluster and stream only the necessary details, reducing CPU and GPU load \cite{overton2024lods, visai2024cinematic}.

In benchmark tests, Nanite-enabled scenes render at an average of 90 FPS on an NVIDIA RTX 3090, compared to 50 FPS in traditional high-polygon rendering methods. Additionally, memory usage is reduced by up to 40\% due to virtualized geometry compression.

\begin{itemize}
    \item \textit{Classification:} Virtualized Micropolygon Geometry System
    \item \textit{Core Principle:} Adaptive Level-of-Detail (LOD) and Real-Time Mesh Streaming
    \item \textit{Algorithm:} Clustered Hierarchical Level-of-Detail (CHLOD)
\end{itemize}

The practical implications of Nanite extend far beyond gaming, offering notable advantages across various industries \cite{wass2024realistic}. For game developers, it enables the creation of highly detailed environments that immerse players without compromising performance. Open-world games, in particular, benefit from the seamless rendering of intricate terrains and structures \cite{li2024real}. Beyond gaming, Nanite proves invaluable in architectural visualization, allowing clients to experience interactive walkthroughs of complex building designs with a realistic sense of scale and fine detail. This versatility highlights Nanite's potential to enhance user experiences in both entertainment and professional applications.

\subsection{Lumen Global Illumination}

Lumen is an advanced real-time lighting system that substantially improves how lighting interacts within virtual environments. Unlike conventional methods that rely on pre-baked lighting data, Lumen provides real-time dynamic lighting capable of adapting to environmental changes instantly. This means that lighting sources, reflections, and shadows respond seamlessly to alterations, whether it’s a shift in a light source, the movement of objects, or changes in the time of day. This adaptability makes Lumen a powerful tool for creating realistic and immersive virtual experiences \cite{epicgames_lumen_2025}.

Lumen achieves 80 FPS on high-end GPUs in a complex urban scene, whereas traditional baked lighting maintains 100+ FPS but lacks real-time adaptability. Dynamic light propagation on mid-range GPUs can introduce a 15--20\% performance cost, making hardware optimization crucial.

This dynamic adaptability is crucial for crafting immersive storytelling experiences, where lighting is central in setting the mood and directing the viewer's focus. Lumen's ability to handle complex lighting scenarios, such as multiple light sources, diffuse reflections, and indirect illumination, further enhances its value. For instance, it accurately replicates the play of sunlight in natural environments, capturing subtle details like light filtering through trees and bouncing off the ground in VR simulations. These capabilities significantly contribute to creating realistic and emotionally engaging virtual worlds \cite{tan2024mastering}. Moreover, Lumen's compatibility with ray tracing technology enhances its functionality, creating cinematic-quality visuals with minimal computational overhead \cite{visai2024cinematic}.

Virtual production is the most salient of Lumen's practical applications. Filmmakers are allowed to immediately try the results of scene lighting, thus giving complete control over every part of scene composition. Lumen's application in game development enables the simulation of day-night cycles and atmospheric sensation, enhancing the immersion into storytelling and gameplay. Moreover, Lumen speeds up the creation process by reducing pre-rendered assets and expanding iterative development.

\begin{itemize}
    \item \textit{Classification:} Dynamic Global Illumination System
    \item \textit{Core Principle:} Real-Time Ray Tracing and Light Propagation Volumes
    \item \textit{Algorithm:} Signed Distance Fields (SDF) and Ray-Marched Reflections
\end{itemize}

To fully appreciate Lumen's strengths, comparing its capabilities with traditional global illumination methods, such as ray tracing and baked lighting, is helpful. These comparisons highlight key differences in adaptability, performance efficiency, visual fidelity, and typical use cases. Table \ref{table:lumen_comparison} provides a clear overview of these distinctions, illustrating how Lumen stands out as a versatile and performance-optimized solution for dynamic lighting in virtual environments.

\begin{table*}[h!]
\centering
\caption{Comparative Analysis of Lumen, Ray Tracing, and Baked Lighting Techniques}
\label{table:lumen_comparison}
\renewcommand{\arraystretch}{1.2} 
\setlength{\tabcolsep}{2pt} 
\begin{tabular}{|>{\raggedright\arraybackslash}p{2.2cm}|
                >{\raggedright\arraybackslash}p{5.2cm}|
                >{\raggedright\arraybackslash}p{5.2cm}|
                >{\raggedright\arraybackslash}p{5.2cm}|} 
\hline
\textbf{Feature}               & \textbf{Lumen}                        & \textbf{Ray Tracing (RTX)}            & \textbf{Baked Lighting}            \\ \hline
\textbf{Dynamic Adaptability}  & Fully dynamic; react in real time to changes in lighting and geometry. & Real-time but computationally intensive; requires high-end GPUs. & Static; pre-computed lighting that does not adapt to changes in real-time. \\ \hline
\textbf{Performance Efficiency} & Optimized for real time applications; suitable for interactive environments. & High computational demands; often requires hardware acceleration. & High preprocessing time but minimal runtime costs. \\ \hline
\textbf{Visual Quality}        & High-quality indirect lighting, diffuse reflections, and soft shadows. & Cinematic-quality reflections, transparency, and caustics. & Limited to pre-defined lighting scenarios; lacks dynamic realism. \\ \hline
\textbf{Primary Use Cases}     & Gaming, VR, and virtual production with dynamic environments. & Film production and high-end visuals, where interactivity is less critical. & Mobile games and static scenes, where computational overhead is a concern. \\ \hline
\end{tabular}
\end{table*}

\subsection{Virtual Shadow Maps}

Virtual Shadow Maps (VSM) in Unreal Engine significantly advance real-time shadow rendering, delivering sharp, detailed shadows that enhance visual realism. Unlike older methods, VSM dynamically adjusts shadow resolution, prioritizing fine details close to the viewer while optimizing performance for distant elements \cite{pisarcik2024render}. This dynamic adaptability allows VSM to handle complex lighting scenes with multiple dynamic light sources interacting with intricate geometries, making it indispensable for open-world video games and virtual reality. For instance, dynamic shadows cast by moving objects or avatars within a virtual forest environment enhance spatial interaction and immersion.

\begin{itemize}
    \item \textit{Classification:} Advanced Shadow Mapping System
    \item \textit{Core Principle:} Adaptive Shadow Map Partitioning
    \item \textit{Algorithm:} Hierarchical Shadow Cascading
\end{itemize}

Compared to conventional cascaded shadow maps, VSM reduces memory consumption by 30\% while delivering higher-resolution soft shadows. In stress tests with dynamic lighting, VSM maintains stable frame rates above 75 FPS on RTX 3080, making it ideal for large-scale environments.

We can see that VSM poses challenges, such as increased memory and computational demands, which can restrict its use on lower-end systems. However, when integrated with Unreal Engine's complementary technologies, like Lumen and Nanite, VSM becomes part of a cohesive rendering pipeline that achieves high-quality visuals without compromising performance. This synergy empowers developers to create shadow-rich, immersive environments, significantly enhancing storytelling and user engagement \cite{epicgames_virtual_shadow_maps_2025}.

\subsection{MetaHuman Creator}

The MetaHuman Creator provides a powerful toolset for creating lifelike models with ease \cite{epicgames_metahuman_2025}. When combined with Unreal Engine's advanced animation systems, such as \textit{motion warping}—which dynamically adjusts animations to match environmental constraints or interactions \cite{motionwarping2022}—and \textit{inverse kinematics (IK)}—which calculates joint movements to ensure natural character interactions with their surroundings \cite{inversekinematics2022}—it contributes meaningfully to character-driven storytelling. These features allow developers to create dynamic, believable characters capable of adapting their movements naturally within interactive environments, enriching emotional engagement with users.

The seamless integration of MetaHuman Creator with motion warping and IK enables characters to dynamically change postures, gesticulate naturally, and interact with objects in their environment, enhancing narrative depth and realism. This is incredibly impactful in VR storytelling, where close-up interactions demand high levels of visual and behavioral authenticity \cite{venter2022unreal}.

\begin{itemize}
    \item \textit{Classification:} Procedural Character Generation System
    \item \textit{Core Principle:} Facial Rigging and Motion Capture
    \item \textit{Algorithm:} Blend Shape Interpolation and Inverse Kinematics
\end{itemize}

Beyond gaming, the MetaHuman Creator has found extensive applications across industries. In virtual production, filmmakers use MetaHuman avatars as digital doubles, minimizing reliance on physical locations and enabling smooth integration with computer-generated imagery (CGI) \cite{maraffi2024metahuman}. Lifelike characters improve engagement in medical simulations, language learning, and other interactive scenarios in education and training. Controlled studies show that realistic avatars enhance user empathy, improve recall, and boost engagement in health and training simulations \cite{rattani2020effectiveness}. 

MetaHuman Creator enables these effects through high-fidelity facial animation and motion capture integration, facilitating emotionally rich and interactive narratives even for small teams. Its user-friendly design further lowers barriers for independent creators, fostering diversity in storytelling and expanding opportunities for character-driven narratives.

While Unreal Engine's core technologies offer technical advancements, their impact on immersive storytelling extends beyond visual fidelity. Table \ref{table:core_features_storytelling} summarizes how each technology directly addresses challenges or creates opportunities in immersive storytelling, going beyond standard technical descriptions.

\begin{table*}[h!]
\centering
\caption{Impact of Unreal Engine 5 Core Features on Immersive Storytelling}
\label{table:core_features_storytelling}
\renewcommand{\arraystretch}{1.2} 
\setlength{\tabcolsep}{2pt} 
\begin{tabular}{|p{2.5cm}|p{7.7cm}|p{7.7cm}|}
\hline
\textbf{Technology} & \textbf{Challenges Addressed} & \textbf{Opportunities for Storytelling} \\ \hline
\textbf{Nanite Virtualized Geometry} & 
Reduces performance issues caused by high-poly assets in open-world environments, eliminating the need for manually created LODs. & 
Allows for cinematic-level detail in interactive environments without sacrificing real-time performance, enabling richer visual storytelling. \\ \hline
\textbf{Lumen Global Illumination} & 
Overcomes the limitation of static lighting in dynamic environments, ensuring real-time global illumination without pre-baked lighting. & 
Enables dynamic mood shifts and lighting-based storytelling techniques, enhancing dramatic tension and environmental storytelling. \\ \hline
\textbf{Virtual Shadow Maps (VSM)} & 
Addresses shadow artifacts and flickering in large-scale scenes, ensuring soft, high-resolution shadows. & 
Improves depth perception and atmospheric immersion, reinforcing visual cues that guide narrative focus. \\ \hline
\textbf{MetaHuman Creator} & 
Removes technical barriers to high-fidelity character creation, previously requiring complex rigging and facial animation workflows. & 
It enables emotionally expressive characters, which is crucial for interactive narratives and virtual performances. \\ \hline
\end{tabular}
\end{table*}

\section{Unreal Engine for Immersive Storytelling}\label{S}

Intending to enable artists to create immersive storytelling experiences, Unreal Engine offers advanced tools and features that combine both interactive design and conventional narrative techniques. Developers can create visually appealing and compelling narratives with Unreal Engine's dynamic lighting, real-time rendering, and robust character generation tools.

\subsection{Intersection of Immersive Storytelling and Game Design}

Unreal Engine integrates storytelling with game interaction, creating new opportunities to deliver immersive narratives. By combining cinematic tools such as \textit{Sequencer} and \textit{Blueprints}, developers can craft stories that fluidly merge with game mechanics, providing a dynamic and engaging user experience \cite{epicgames_sequencer_blueprint_2025}. 

\subsubsection{Sequencer: Crafting Immersive Cinematics}

Sequencer, Unreal Engine's cinematic editor, empowers developers to create and play high-quality cinematic sequences in real-time \cite{epicgamessequencer2025}. It supports a wide range of features, including camera angles, transitions, animations, and visual effects, enabling the seamless integration of cutscenes into gameplay without disrupting the user experience. 

A notable example of its use is in the development of \textit{Hellblade: Senua's Sacrifice}\footnote{\textit{Hellblade: Senua's Sacrifice} is an award-winning adventure game for its innovative storytelling and its sensitive representation of mental health \cite{soc14090170}.}. The Sequencer was instrumental in merging the game’s physiological storytelling with real-world gameplay mechanics \cite{fordham2019framing}. This integration ensured smooth transitions between emotionally charged cutscenes and interactive sequences, deepening players' emotional connection to the protagonist’s journey. The Sequencer’s capability to incorporate motion capture data and synchronize audio with visual effects further expanded its potential for crafting immersive narratives.

\subsubsection{Blueprints: Simplifying Interactive Design}

Blueprints, Unreal Engine's visual scripting system, enables developers to create gameplay mechanics without writing code. Using a node-based interface, developers can build logic by connecting nodes and creating graphs, democratizing access to complex game design previously reserved for programmers \cite{EpicGamesBlueprints, sewell2015blueprints}. This system is especially valuable for interdisciplinary teams, allowing artists and writers to contribute directly to gameplay design and narrative development without relying entirely on programming expertise.

Blueprints are equally effective for both indie and AAA\footnote{AAA refers to high-budget, high-profile games developed by large studios with significant resources, offering high-grade graphics.} projects. They are widely used for creating environmental puzzles, interactive story triggers, and branching dialogue systems \cite{boyd2017reinforcement}. Blueprints foster creativity and innovation by empowering diverse teams to experiment and iterate, even within smaller teams.

Sequencer and Blueprints embody Unreal Engine's commitment to blending storytelling with interactivity. These tools not only streamline the development process but also improve the overall quality of the final product. Leveraging these innovations, creators can design immersive narratives that deeply engage players, enriching the overall gaming experience \cite{epicgames_sequencer_blueprint_2025}.

\subsection{Facilitating Immersive Narratives in VR}

Unreal Engine has become a cornerstone for crafting immersive VR storytelling experiences by leveraging real-time rendering, spatial audio, and dynamic lighting. These tools are particularly crucial in VR, where creating a sense of presence and interactivity is essential. Real-time rendering enables virtual environments to respond seamlessly to user interactions, preserving visual fidelity and maintaining immersion as users navigate intricate and detailed spaces \cite{epicgames_real_time_rendering_intro}. Spatial audio further enhances this experience by generating ambient soundscapes that adapt to the player's movements, aligning sound with their perspective to provide intricacies and emotional depth \cite{broderick2018importance}. Meanwhile, dynamic lighting enhances realism by reacting to environmental changes, supporting dramatic storytelling through mood shifts or transitions in time \cite{EpicGamesLighting2023}.

\begin{figure}[h]
    \centering
\includegraphics[width=0.45\textwidth]{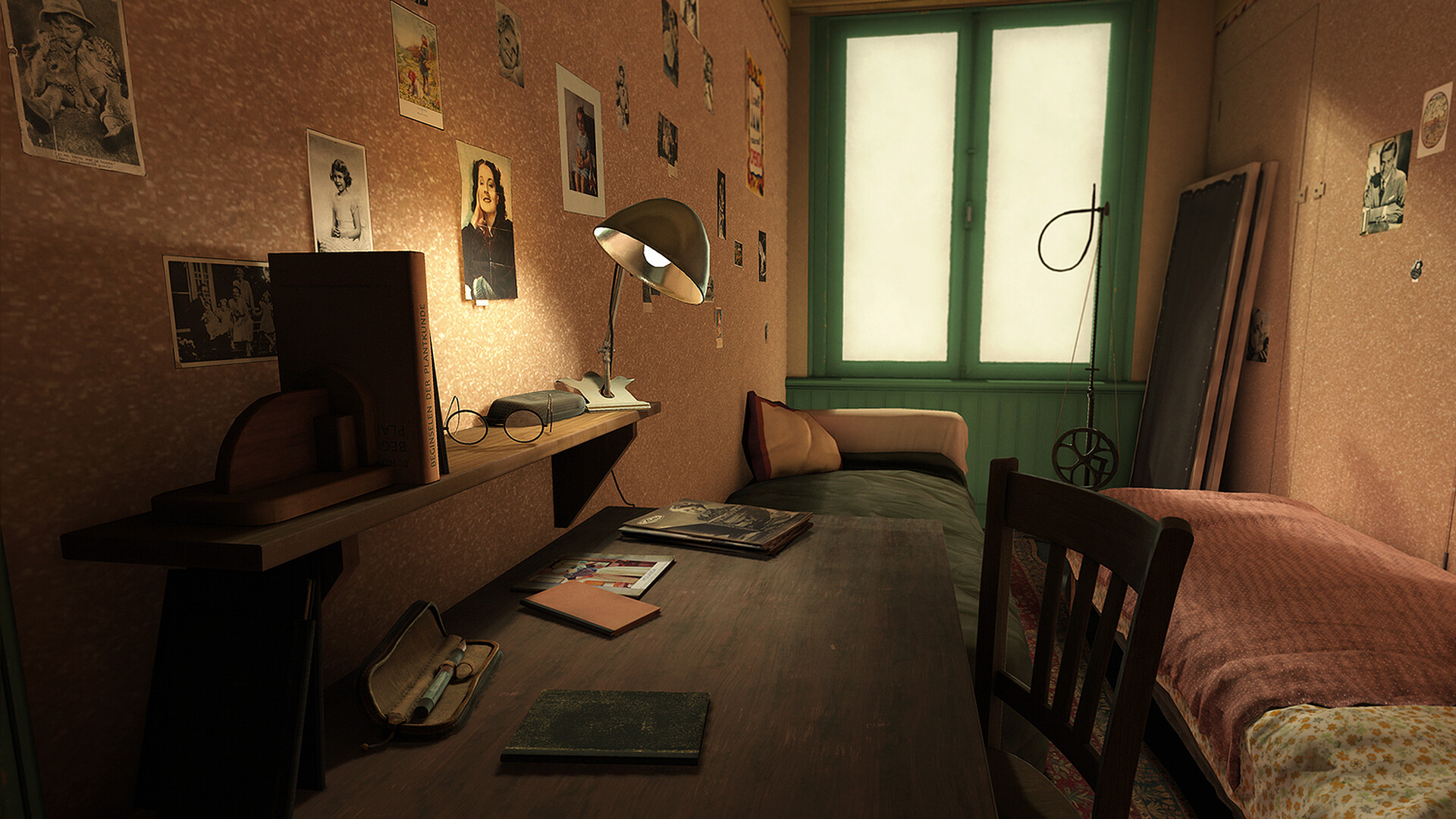}
    \caption{The annex environment recreated in \textit{The Anne Frank House VR }\cite{AnneFrankHouseVR}.}
    \label{fig:anne_frank_Figure}
\end{figure}

An exemplary application of these capabilities  House VR, depicted in Figure \ref{fig:anne_frank_Figure}. This project utilizes Unreal Engine to faithfully recreate the secret annex where Anne Frank and her family hid during World War II, achieving exceptional historical accuracy \cite{hartmann2013anne}. Users can explore this historically significant space, accompanied by excerpts from Anne's diary, with visually precise environments and emotionally resonant sound design immersing them in the narrative. This combination offers an educational and profoundly moving experience that underscores the potential of Unreal Engine in VR storytelling.

Similarly, projects like Tilt Brush harness Unreal Engine's advanced tools to enable interactive artistic expression \cite{haeyen2021use}. This application allows users to paint in immersive 3D spaces enriched by dynamic lighting and responsive textures, blending creativity with a deeply engaging and immersive experience.

Unreal Engine's versatility extends beyond artistic endeavors, encompassing diverse VR applications in gaming, education, and cultural preservation. Its combination of technical excellence and creative flexibility empowers developers to redefine the boundaries of interactive storytelling, crafting narratives that captivate and emotionally resonate with audiences. As VR technology evolves, Unreal Engine remains at the forefront, offering the tools and capabilities to create transformative and unforgettable experiences \cite{broderick2018importance}.

\section{Applications of Unreal Engine}\label{A}

Unreal Engine empowers creators to design immersive and visually stunning experiences through its extensive suite of tools. Its adaptability and scalability make it a versatile platform, ideal for various applications across various industries \cite{mack2019unreal}.
This section explores the transformative impact of Unreal Engine across multiple domains, highlighting how it redefines workflows, unleashes creative potential, and addresses complex challenges.

\subsection{Game Development}

\begin{figure}[h]
    \centering
\includegraphics[width=0.45\textwidth]{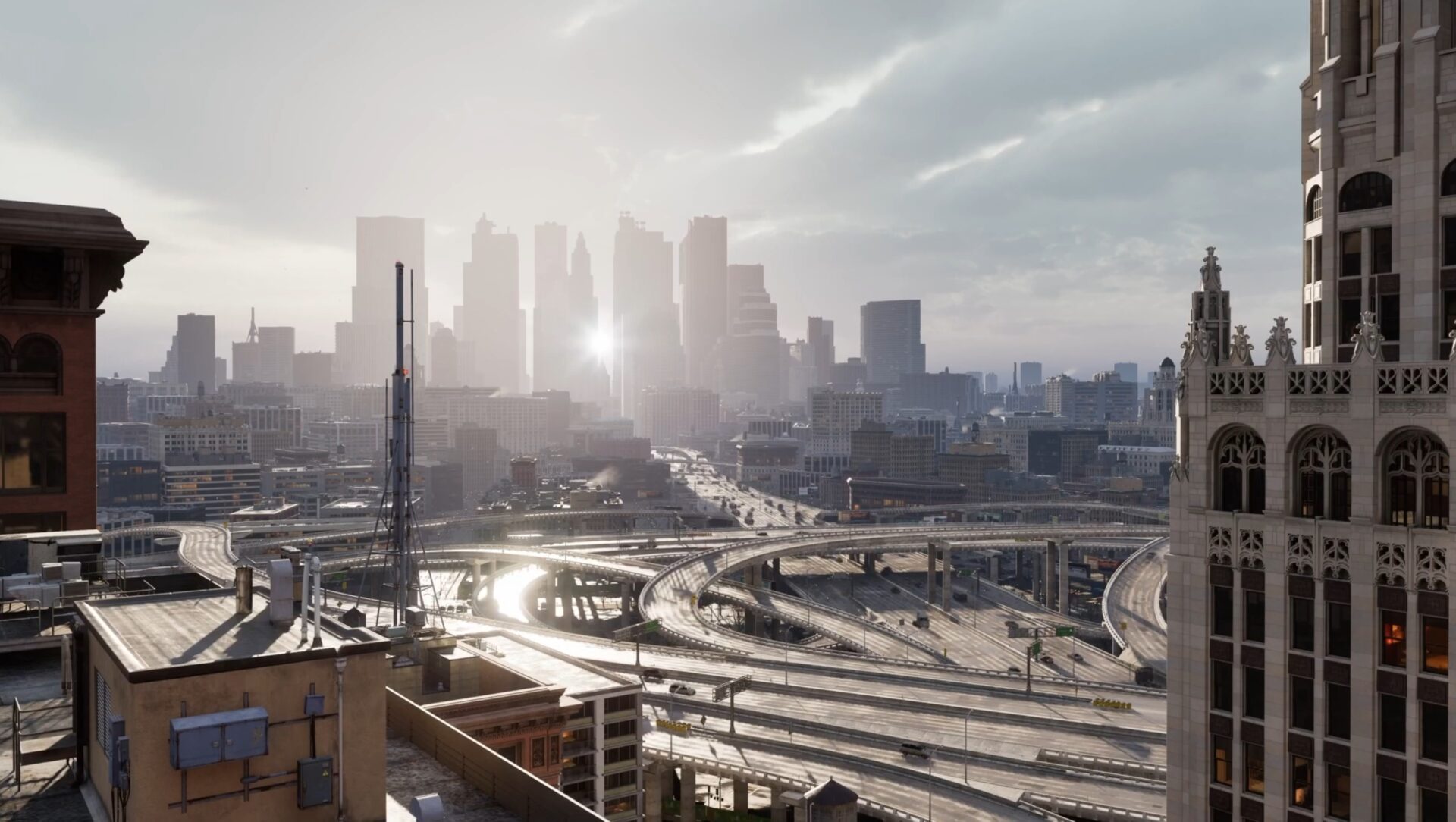}
    \caption{Screenshot from \textit{The Matrix Awakens} showcasing Unreal Engine's realistic rendering capabilities \cite{UE5CityVideo}.}
    \label{fig:matrix}
\end{figure}

Unreal Engine has powered the creation of numerous games, including \textit{The Matrix Awakens} \cite{el2022unreal}, a tech demo that vividly showcases its ability to render expansive open worlds with remarkable realism, as illustrated in Figure \ref{fig:matrix}. 
Similarly, \textit{CD Projekt Red}, the studio behind acclaimed titles like \textit{The Witcher} and \textit{Cyberpunk 2077}, has adopted Unreal Engine for the next installment of \textit{The Witcher} series. This decision underscores the engine's scalability and suitability for crafting immersive, large-scale, open-world designs.

Another notable project is \textit{Senua's Saga: Hellblade II}, which utilizes Unreal Engine's MetaHuman Creator to develop hyper-realistic characters. This tool captures subtle facial expressions and movements, enabling the creation of emotionally charged narratives that raise awareness of mental health issues through transformative play \cite{austin2021hardest, fusdahl2019vulnerability}. Such games exemplify how Unreal Engine is shaping the future of AAA game development.

To optimize performance, developers leverage various strategies within Unreal Engine. For instance, streaming virtual textures efficiently handles large volumes of assets with minimal memory usage. At the same time, the integration of LOD adjustments, combined with Nanite, ensures smooth transitions even in complex environments \cite{fritsch2004visualisation}. Additionally, effective GPU resource management enables real-time rendering of intricate effects, such as particle systems and dynamic lighting, in expansive game worlds \cite{wang2016real, kato2011timegraph}. 

These game development advancements enhance visual realism and deepen player immersion by ensuring seamless storytelling integration. Creating expansive, dynamic worlds with photorealistic lighting and responsive environments enables developers to craft emotionally engaging narratives that react to player choices and interactions with others \cite{multiplayer}. By reducing technical barriers, Unreal Engine facilitates visually stunning and intensely interactive storytelling.

\subsection{Virtual Production}

Apart from gaming, Unreal Engine is widely recognized for revolutionizing virtual production, transforming how films and television shows are created. By integrating real-time rendering into traditional workflows, Unreal Engine allows filmmakers to dynamically adjust scenes during production. Advanced features of Unreal Engine, such as real-time ray tracing \cite{an2022technology} and volumetric lighting \cite{lama2024future}, enable the creation of highly realistic environments and seamless blending of physical and virtual elements.

% Review 2 Reviewer 2 comment 3
A standout example of Unreal Engine in virtual production is Disney+'s \textit{The Mandalorian}. The series utilized Unreal Engine as the foundation for StageCraft, an advanced virtual production platform that combines LED screen technology with real-time rendering. One of the critical enabling technologies was Lumen, which allowed the lighting on physical actors and props to adapt to the virtual backgrounds displayed on the LED walls in real time. This meant that as directors or cinematographers adjusted virtual lights or changed digital environments on the fly, the lighting conditions in the physical set responded naturally and immediately, preserving visual continuity and eliminating the need for extensive relighting or compositing in post-production. 

\begin{figure}[h]
    \centering
\includegraphics[width=0.45\textwidth]{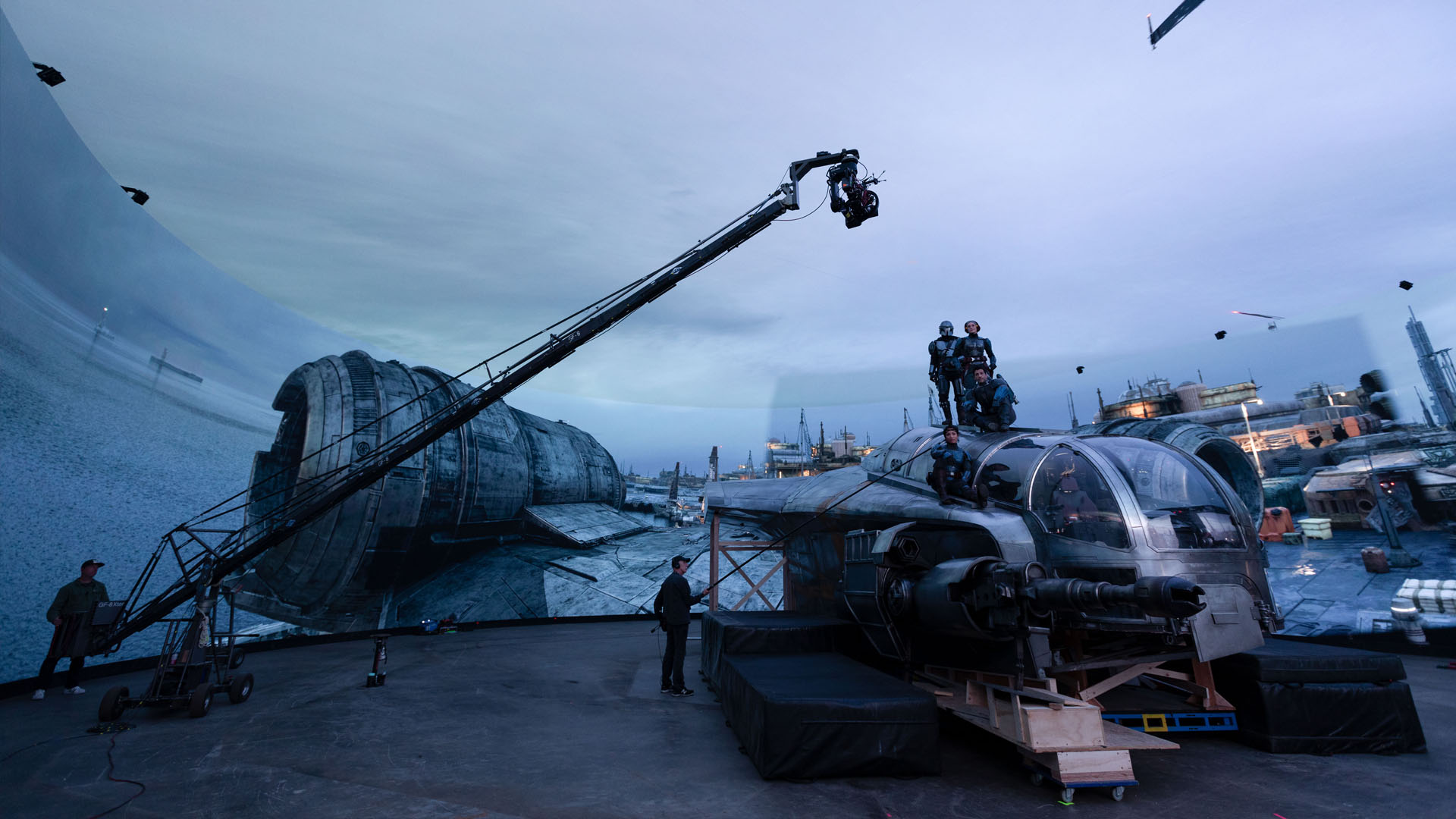}
    \caption{Photo from \textit{The Mandalorian} showing how StageCraft was implemented on scene \cite{mandalorianLEDWall}.}
    \label{fig:mandalorian}
\end{figure}

This flexibility enhanced the creative process and significantly accelerated production workflows by reducing dependency on static lighting setups and pre-rendered content. As illustrated in Figure~\ref{fig:mandalorian}, integrating physical and virtual lighting in real time contributed to the cinematic realism that defines the show.

%--

Similarly, Netflix's \textit{Love, Death and Robots} leveraged Unreal Engine for final rendering, streamlining the animation pipeline \cite{nowak2018modeling}. Utilizing Unreal Engine's advanced rendering capabilities, the production team created photorealistic human characters with precisely choreographed fight sequences and dialogues while accelerating the workflow. This efficiency allowed the team to concentrate on the creative aspects of storytelling, showcasing Unreal Engine's value in animation and visual effects \cite{SonyImageworks2023}.

Features like real-time ray tracing enhance production quality by delivering accurate reflections, shadows, and global illumination, ensuring lifelike lighting that adapts naturally to scene dynamics \cite{peddie2019ray}. In-camera visual effects further synchronize camera movements with rendered environments, enabling directors to capture authentic shots seamlessly, even when blending live-action elements with digital assets \cite{hodgkinson2017lights}. Additionally, Unreal Engine's AI-assisted workflows automate repetitive tasks such as environment detailing and asset placement, significantly speeding up production \cite{sapio2019hands}. 

Virtual production extends beyond efficiency—it transforms storytelling by allowing filmmakers to create dynamic, photorealistic worlds in real-time. This capability enhances narrative fluidity, enabling directors to experiment with lighting, environments, and character positioning in ways that traditional filmmaking does not allow. Through real-time adjustments and in-camera visual effects, Unreal Engine ensures that immersive storytelling remains at the heart of digital content creation.

\subsection{Architectural Visualization and Simulations}

Unreal Engine offers powerful tools for advanced photorealistic rendering and interactive features in architectural visualization. Lumen and Nanite enable the creation of immersive environments that closely replicate real-world conditions. These tools allow users to experience designs as they were built, fostering improved communication and a deeper understanding of project realization \cite{david2022integrating}.

Beyond individual projects, Unreal Engine powers smart-city simulations to model complex scenarios such as traffic flow, infrastructure changes, and environmental impact in real-time \cite{buyukdemircioglu2022development}. These simulations provide a dynamic platform for testing design decisions, enhancing accuracy and efficiency in urban planning and development.

The benefits of Unreal Engine extend beyond visual aesthetics. Accurate lighting simulations help architects demonstrate how natural and artificial light interact with spaces, enabling clients to visualize environments under various conditions. This capability is invaluable for ensuring that lighting meets functional and aesthetic requirements \cite{natephra2017integrating}. 

Interactive walkthroughs further elevate the visualization process by allowing users to explore spaces from multiple perspectives \cite{shiratuddin2002virtual}. Unlike static renderings, these walkthroughs offer a comprehensive understanding of scale, layout, and spatial relationships, instilling greater confidence in design decisions while improving project accessibility \cite{david2022integrating}.

\begin{figure}[h]
    \centering
\includegraphics[width=0.45\textwidth]{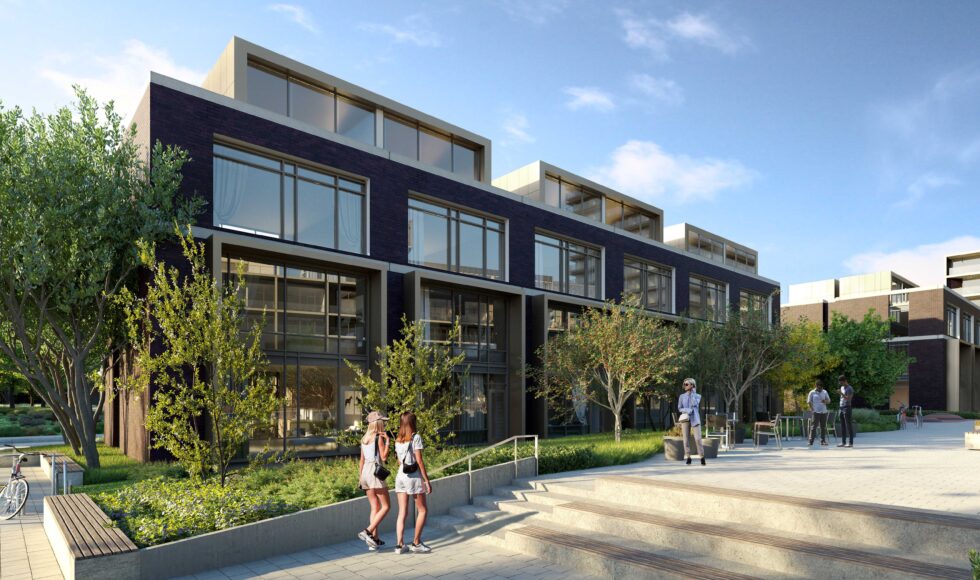}
    \caption{A photorealistic rendering created with Unreal Engine, demonstrating the capabilities of Lumen and Nanite technologies \cite{radicalGalaxy2025}.}
    \label{fig:architecture}
\end{figure}

Unreal Engine has made architectural visualization more practical, accessible, and impactful. From virtual showrooms to urban planning simulations, Unreal Engine simplifies processes for architects, designers, and clients, fostering innovative and efficient design solutions (See Figure~\ref{fig:architecture}).

Architectural visualization leverages many of the same immersive storytelling principles used in virtual production and gaming. Real-time rendered environments, interactive exploration, and dynamic lighting allow users to engage with spaces in a narrative-like experience, where the design unfolds as they move through it. The ability to guide users’ focus through environmental storytelling—using light, texture, and spatial design—demonstrates how Unreal Engine bridges architecture and narrative-driven immersion. By crafting digital experiences that tell the story of a building's function, history, or evolution, Unreal Engine enables architects and designers to create compelling spatial narratives beyond static blueprints or traditional 3D renders.

\subsection{Applications in Cultural Preservation and Social Impact Storytelling}

Unreal Engine has emerged as a powerful tool for cultural preservation and social impact storytelling, enabling audiences to engage with history and humanitarian issues in immersive and meaningful ways \cite{denchev2020application}. By leveraging advanced technologies, developers can craft interactive experiences that bridge the past and present, creating connections that are both impactful and educational \cite{lepouras2004virtual}.

\begin{figure}[h]
    \centering
\includegraphics[width=0.46\textwidth]{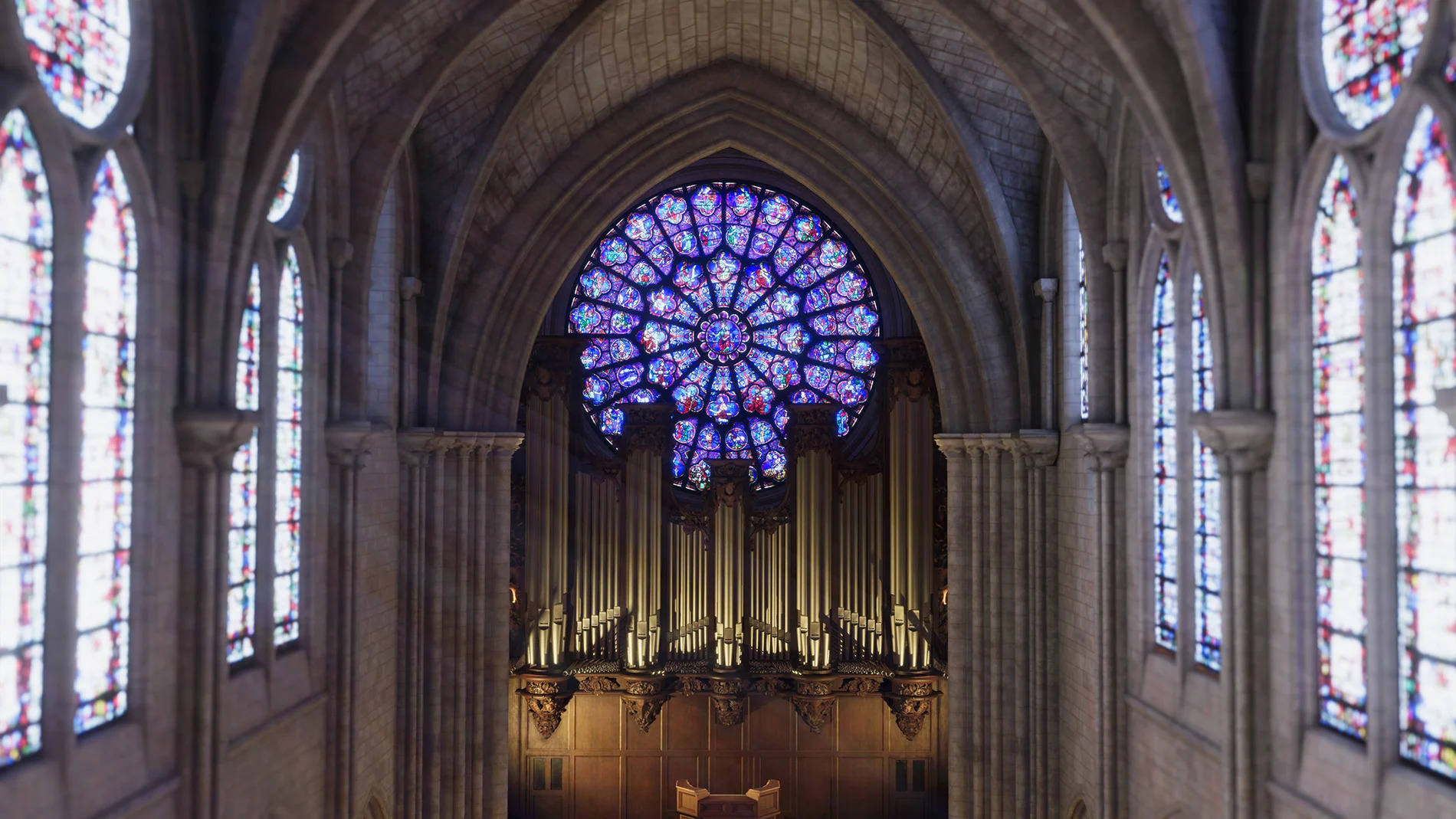}
    \caption{Screenshot from \textit{Eternal Notre-Dame} showing realistic recreation of cathedral \cite{enVolsNotreDame}.}
    \label{fig:notre_dame}
\end{figure}

A striking example is the Eternal Notre-Dame VR (See Figure~\ref{fig:notre_dame}), a project aimed at digitally recreating the iconic cathedral after the devastating 2019 fire \cite{postma2016virtual, comte2024strategies}. Using Unreal Engine, developers meticulously reconstructed the site, offering virtual tours that showcase the cathedral’s interiors and exteriors with a high level of detail. This project authenticates the structure through real-time rendering and realistic textures, granting global audiences access to a cultural heritage site that would otherwise remain inaccessible during restoration. Such initiatives preserve historical landmarks and offer immersive educational opportunities surpassing static media.

Unreal Engine also plays a pivotal role in social advocacy projects. For instance, the VR documentary \textit{Clouds Over Sidra}, developed in collaboration with the United Nations, transports viewers to Jordan’s Za’atari refugee camp \cite{suzuki2022limits}. This experience follows Sidra, a 12-year-old Syrian girl, using Unreal Engine to create a 360-degree immersive environment that portrays her daily life. The project delivers a visually compelling and emotionally resonant narrative, fostering empathy and raising awareness about the challenges faced by refugees \cite{alberghini2020improving}.

These applications highlight the role of Unreal Engine in crafting emotionally resonant experiences that connect audiences with historical and cultural narratives. By leveraging high-fidelity environments and real-time interactivity, these projects go beyond traditional storytelling, allowing users to engage with history in a deeply personal and immersive way \cite{oing2018implementations}. This reinforces the broader theme of how Unreal Engine bridges digital innovation and meaningful storytelling.

\subsection{Emerging Applications}

Unreal Engine extends beyond VR into fields such as extended reality (XR), providing innovative solutions across diverse domains. Platforms like \textit{EngageVR} leverage Unreal Engine to create educational experiences that transport users to historical landmarks or natural habitats, making learning more interactive and engaging. These virtual environments enable students to explore and interact with places they might never physically visit, enhancing educational outcomes \cite{perinpasingam2023exploring}.

The engine's influence also extends to healthcare, where VR rehabilitation programs powered by Unreal Engine assist patients in regaining mobility and functionality. These programs offer interactive and adaptive recovery approaches, demonstrating how XR technology can support therapeutic interventions effectively and integrate immersive solutions into everyday practices \cite{lew2021virtual, al2019use}.

%\begin{figure}[h]
%    \centering
%\includegraphics[width=0.8\textwidth]{picture-9.jpg}
%    \caption{A VR rehabilitation program used for restoration of hand motor functions \cite{ProgramAceVRinHealthcare}.}
%    \label{fig:rehabilitation}
%\end{figure}

Unreal Engine further excels in AI-driven workflows, particularly in procedural content generation. By incorporating machine learning algorithms, creators have greater freedom to design intricate environments with minimal manual effort. This capability is especially beneficial in game development and urban planning, where scalability and efficiency are critical \cite{abu2023new}. Additionally, AI tools within Unreal Engine enhance character behavior modeling, enabling responsive interactions and realistic simulations. For instance, AI-generated crowd behaviors and adaptive Non-Playable Characters (NPCs) add dynamic realism to games and simulations, elevating the overall immersive experience \cite{rodrigues2021shriek}.

These applications highlight Unreal Engine's versatility in combining immersive XR technologies and AI to drive innovation across education, healthcare, and interactive simulations, advancing creative and practical solutions.

While not traditionally associated with storytelling, many of these emerging applications still rely on immersive engagement techniques. AI-driven procedural generation enables dynamic world-building, allowing environments to adapt and evolve in response to user interactions, much like a living narrative. Similarly, educational and rehabilitation applications benefit from interactive storytelling elements, where users progress through experiences shaped by their actions \cite{damianova2025serious}. In XR-based learning, Unreal Engine creates scenarios that function like interactive narratives, guiding users through structured yet responsive experiences that foster more profound understanding and emotional connection. By integrating these storytelling principles, Unreal Engine enhances immersion across disciplines beyond entertainment, reinforcing its role as a transformative tool in digital experiences.

% Review 2 Reviewer 2 comment 6
\section{Techniques for Immersive Narratives}\label{T}

Immersive narratives merge technology with storytelling, creative experiences that place users directly into the story world. VR and interactive environments respond to user actions, fostering a sense of presence. Unreal Engine achieves it by utilizing spatial audio for interactive dialogues, and responsive environmental design engages with multiple senses, making narration more memorable. Blueprints enable interactivity, while MetaHuman Creator and Sequencer evoke emotional engagement through characters and synchronized visuals and sound \cite{jacobson2005caveut}.

\subsection{Spatial Storytelling and Environment Design}

Immersive storytelling relies on interactive elements and thoughtfully crafted environments, forming its foundation. Interactivity transforms audiences from passive observers into active participants, allowing them to engage with the narrative on a deeper level. Features such as branching dialogues adapt dynamically to user choices, enabling players to shape the story’s progression according to their preferences. Unreal Engine facilitates this level of interactivity through tools like Blueprints \cite{yao2022design}.

Equally important is environmental design, which enhances immersion by providing the context and atmosphere for the story. Spatial audio plays a key role in guiding users with auditory cues highlighting important moments or locations within the narrative. Unreal Engine’s advanced audio processing capabilities ensure precise sound specialization, creating highly realistic auditory experiences. Meanwhile, Lumen’s dynamic lighting brings additional depth by highlighting focal points and adjusting the mood of a scene as the story unfolds, reinforcing both emotional impact and narrative depth \cite{wiesing2020accuracy}.

\begin{figure}[h]
    \centering
\includegraphics[width=0.45\textwidth]{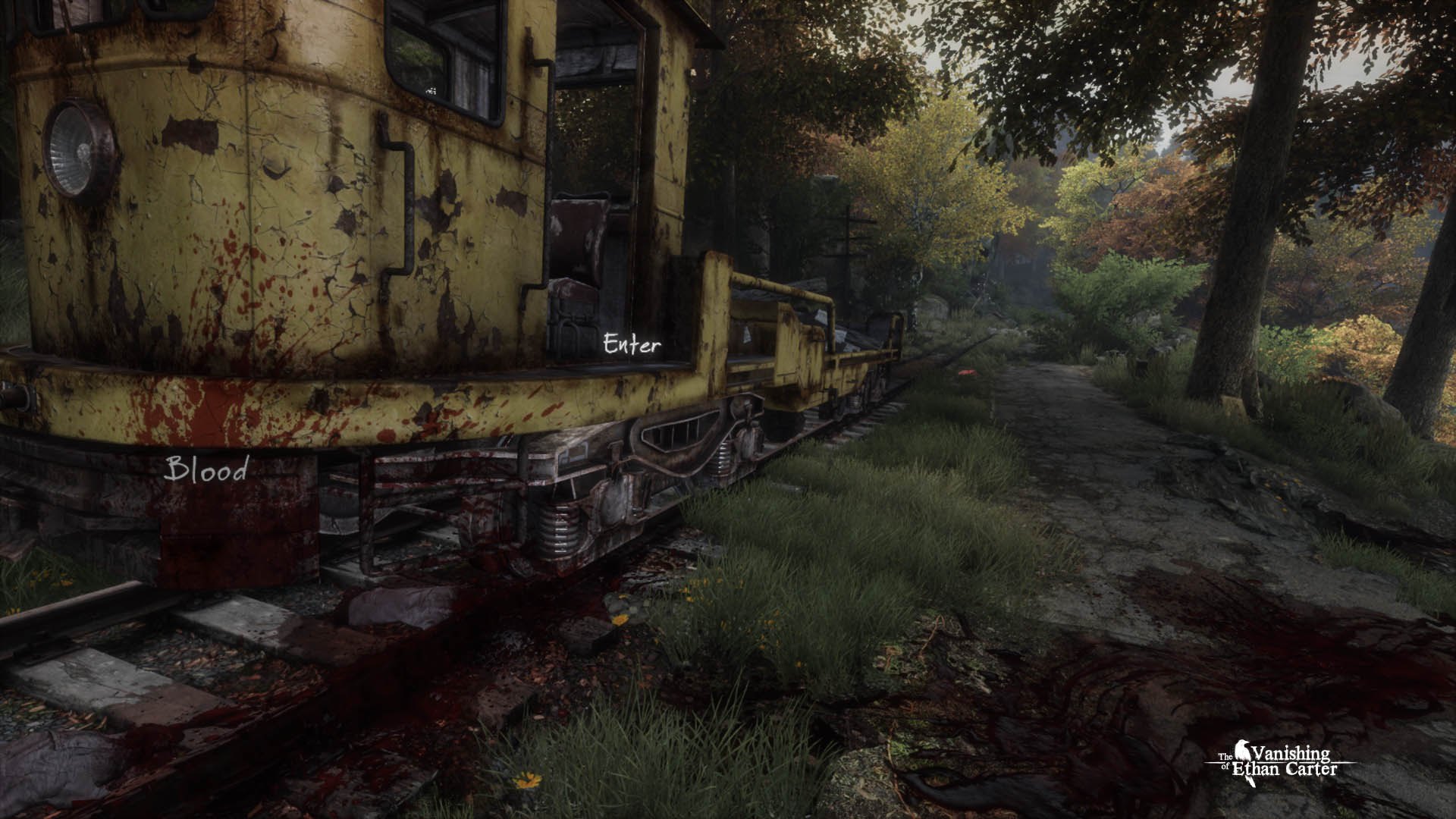}
    \caption{Screenshot from \textit{The Vanishing of Ethan Carter} showing how clues are telling the story with no explanation needed \cite{VanishingOfEthanCarter}.}
    \label{fig:ethan}
\end{figure}

A compelling example of these principles in action is \textit{The Vanishing of Ethan Carter}, a game that uses environmental storytelling to unravel a mystery \cite{statham2020use}. Instead of relying on traditional gameplay mechanics like quest markers or detailed objectives, the game emphasizes exploration and discovery. Players uncover the story organically through interactions with the environment, as illustrated in Figure~\ref{fig:ethan}. Spatial audio elements, such as the rustling of leaves or distant whispers, subtly guide exploration without overt direction. This project demonstrates how Unreal Engine’s tools can create responsive and immersive narrative experiences, offering a fresh approach that transcends traditional game design.

\subsection{Character Interaction and Dialogue Systems}

MetaHuman Creator and Sequencer evoke emotional engagement through characters and synchronized visuals and sound. These tools support lifelike performances and allow for branching narratives and dynamic responses.

Interactive characters are enhanced through procedural animation systems, such as motion warping and inverse kinematics. These systems adjust character posture and limb movement in real-time, ensuring naturalistic behavior during player interaction. Dialogues can be implemented using Blueprint-based logic or third-party AI tools, enabling NPCs to respond meaningfully to player behavior.

\subsection{Emotional Immersion in VR}

Emotional engagement is a cornerstone of immersive storytelling, and Unreal Engine provides an array of tools to bring it to life. Central to this is the MetaHuman Creator, which allows developers to create lifelike characters with expressive facial animations and realistic body movements. These characters can convey nuanced emotions, such as joy, fear, or sadness, helping users connect more deeply with the narrative\cite{heydarian2015immersive}.

Lighting and color also play a pivotal role in shaping emotional storytelling. Unreal Engine’s Lumen technology dynamically adapts to changes in mood or atmosphere in real time. For example, transitioning from warm, golden tones to cold, desaturated lighting can effectively signal a narrative shift, evoke tension, or create unease, aligning the visual environment with the story’s emotional arc\cite{isbister2016games}.

Equally important is sound design, which enhances emotional impact through spatial audio. This feature allows developers to position sounds within a 3D space, creating a sense of proximity and directionality. Subtle auditory cues—such as a distant melody or the sudden sound of footsteps—can heighten suspense and draw users further into the experience.

\begin{figure}[h]
    \centering
\includegraphics[width=0.45\textwidth]{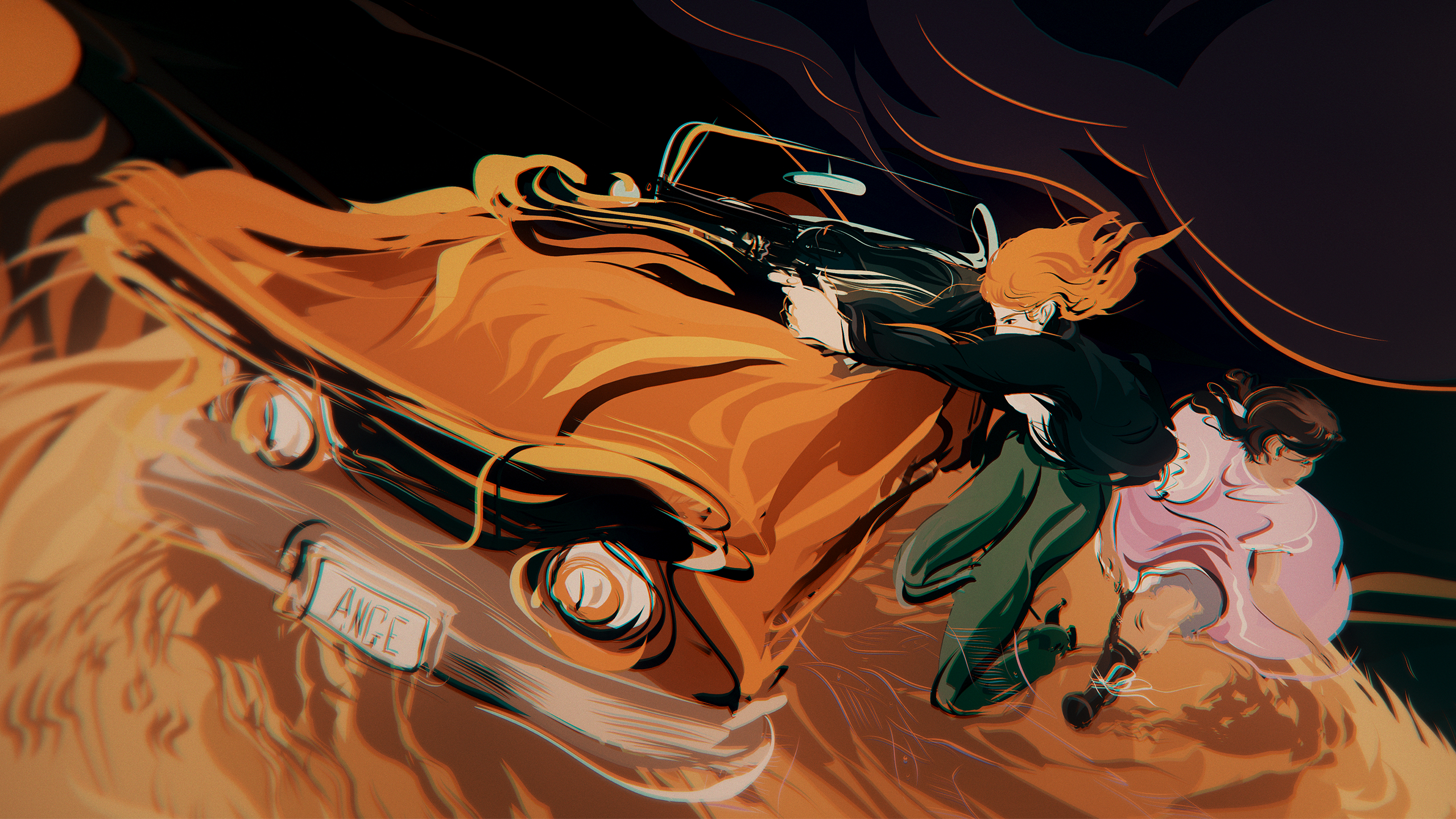}
    \caption{Screenshot from \textit{Dear Angelica} showcasing the transformation of a family road trip into a “Thelma and Louise” \cite{VarietyDearAngelica2017}.}
    \label{fig:angelica}
\end{figure}

Projects like \textit{Dear Angelica} \cite{oculus2017dear, hukerikar2023analyzing} and \textit{Project Syria} \cite{garsuta2021virtual, arar2024virtual} exemplify how these elements work in harmony. By combining advanced sound design, dynamic lighting, and expressive character animation, these projects craft emotionally resonant VR narratives that leave a lasting impression on audiences, showcasing the power of Unreal Engine in delivering impactful storytelling experiences (see Figure~\ref{fig:angelica}).

%--
\section{Comparison with Other Engines}\label{COM}

Unreal Engine, Unity, and CryEngine are among the most popular game engines, each offering unique strengths and weaknesses, particularly for VR projects. Their varying features, performance, and ease of use make them suitable for different types of development \cite{barczak2019comparative}.

Unreal Engine stands out for its high visual fidelity, making it the go-to choice for projects requiring realistic rendering and dynamic lighting. Advanced features like Nanite and Lumen enable high-definition graphics and real-time illumination with minimal optimization efforts. These capabilities make Unreal Engine ideal for AAA games, virtual production, and VR applications where realism is critical. Unreal Engine's Blueprints scripting system also provides an intuitive interface for defining game logic, allowing collaboration between programmers and non-programmers \cite{multiplayer}. However, these advanced features come with the trade-off of high hardware demands, which can pose challenges for smaller teams.

Unity, in contrast, is renowned for its flexibility and accessibility, particularly favored by indie game developers \cite{goldstone2009unity}. Its lightweight architecture and extensive asset store simplify rapid prototyping and development on limited budgets. While Unity supports VR and AR applications, its visual capabilities fall short of Unreal Engine's, especially in rendering high-end graphics and complex lighting \cite{jerald2014developing}. Unity's C\# scripting environment is straightforward and robust, and its performance optimization tools, such as adaptive resolution, make it an excellent choice for scalable projects across diverse platforms \cite{aversa2019unity}.

CryEngine excels in rendering outdoor and natural environments, with advanced terrain and vegetation systems that are unmatched for creating expansive landscapes \cite{kaplanyan2009light}. These features make it particularly well-suited for first-person shooters and simulations requiring high-quality visuals, including VR applications \cite{juarez2010implementing}. However, CryEngine’s steep learning curve and limited community support make it less accessible to beginners and smaller teams. Its sparse documentation and third-party resources further add to its challenges. However, we selected Unreal Engine based on the benchmark comparison made by previous literature and developers that position Unreal Engine as the best tool nowadays in immersive storytelling. 

\subsection{Performance Benchmark Comparison}

Table~\ref{tab:engine_comparison} presents performance benchmarks for VR applications developed in each engine, based on publicly available data from comparable projects run on similar hardware (RTX 3080, 32GB RAM, 1440p resolution) \cite{bao2024comparative, sapio2022developing, gundlach2014mastering, hocking2022unity}. 

\begin{table*}[h]
\centering
\renewcommand{\arraystretch}{0.9} % Reduced row height
\setlength{\tabcolsep}{2pt} % Reduced column spacing
\caption{Comparison of Unreal Engine, Unity, and CryEngine for Storytelling in VR.}
\label{tab:engine_comparison}
\begin{tabular}{|p{2.2cm}|p{5.2cm}|p{5cm}|p{5cm}|}
\hline
\textbf{Feature} & \textbf{Unreal Engine} & \textbf{Unity} & \textbf{CryEngine} \\ \hline
\textbf{Visual Fidelity} & 
Leading photorealism with Nanite and Lumen. Ideal for AAA games. & 
Good quality, less advanced. Great for mobile or indie games. & 
Top rendering for natural landscapes. Ideal for simulations and shooters. \\ \hline
\textbf{Ease of Use} & 
Blueprints' visual scripting for non-programmers. & 
User-friendly C\# scripting and asset store. Perfect for small teams. & 
Steep learning curve, limited resources. Requires expertise. \\ \hline
\textbf{VR/AR Support} & 
Comprehensive VR tools for interactivity and realism. & 
Strong mobile and cross-platform AR/VR support. & 
Good VR focuses on visuals but is less versatile than Unreal or Unity. \\ \hline
\textbf{Community and Resources} & 
Large active community, extensive documentation. & 
Vast support, marketplace, and learning materials. & 
Smaller community, fewer tools, limited documentation. \\ \hline
\textbf{Best Use Cases} & 
AAA games, virtual production, cinematic storytelling, high-end VR. & 
Mobile games, indie projects, scalable AR/VR, educational tools. & 
Simulations, shooters, and realistic outdoor environments. \\ \hline
\end{tabular}
\end{table*}

Unreal Engine demonstrates superior visual fidelity and specialized tools tailored for storytelling and VR, mainly due to features like Nanite’s auto-LOD streaming, which minimizes draw calls and enhances rendering performance, albeit at the cost of slightly higher RAM usage and larger build sizes. Unity, by contrast, prioritizes flexibility and accessibility, making it ideal for indie developers and cross-platform projects, though it often requires more manual optimization. CryEngine offers a middle ground in performance, particularly for natural environment simulations, but it falls short in usability, community support, and VR-specific tooling compared to its competitors.

\subsection{Case Study Comparison: VR Storytelling Project}

Table~\ref{tab:comparison} compares how each engine performed in similar narrative-driven VR projects to further illustrate differences. The comparison highlights their strengths and weaknesses in key areas such as graphics rendering, usability, performance optimization, and cross-platform support \cite{vohera2021game, goldstone2009unity, soni2024merits}. 

\begin{table*}[h!]
\centering
\caption{Comparative Analysis of Unreal Engine and Unity in Technical Level}
\label{tab:comparison}
\begin{tabular}{|p{2.2cm}|p{7.3cm}|p{7.3cm}|}
\hline
\textbf{Feature} & \textbf{Unreal Engine} & \textbf{Unity} \\ \hline
\textbf{Graphics Rendering} & Advanced rendering with Lumen, Nanite, and PBR for high visual fidelity. Supports global illumination, HDR, and ray tracing. & It is less performant for high-end rendering, but offers flexibility for simpler projects. \\ \hline
\textbf{Usability} & Blueprints simplify some tasks but require advanced knowledge for optimal use. & Beginner-friendly with an intuitive interface and vast documentation. Occasional challenges with error messages and UI. \\ \hline
\textbf{Asset Store} & Fab asset store integrates multiple platforms (e.g., Sketchfab, Quixel). Launched in 2024 and rapidly growing. & Extensive Unity Asset Store has many assets, including 3D/2D models, templates, AI tools, and more. \\ \hline
\textbf{Hardware Requirements} & Higher requirements for optimal performance, including RTX 2000 series or better for ray tracing and 32 GB of RAM. & More lenient requirements. It can run on lower-end devices with minimal hardware (e.g., 1 GB RAM for mobile). \\ \hline
\textbf{Performance Benchmarks} & 80-120 FPS (Nanite \& Lumen on RTX 3090), optimized memory usage. & 90-140 FPS in lightweight scenes, varies with HDRP/URP. \\ \hline
\textbf{Cross-Platform Support} & Supports major platforms but with a focus on high-performance devices. & Extensive cross-platform support for 17 platforms, including PCs, consoles, mobile, and WebGL. \\ \hline
\textbf{Performance Optimization} & Nanite reduces computational complexity, enabling efficient rendering of high-polygon models. & Flexible optimization options for mid-range and low-end devices through URP and custom scripts. \\ \hline
\textbf{Target Audience} & Suited for projects requiring cutting-edge visuals and developers with advanced technical skills. & Ideal for beginners, small teams, and projects prioritizing accessibility and rapid development. \\ \hline
\textbf{Customization and Flexibility} & Extensive options for customization, but require significant expertise in C++ and Blueprints. & Highly flexible, with support for custom scripts using C\# and extensive third-party integrations. \\ \hline
\end{tabular}
\end{table*}

These case studies show Unreal Engine’s advantage in emotional storytelling and visual expressiveness, especially with tools like Sequencer and MetaHuman. Unity excels in lightweight, mobile-first storytelling, while CryEngine remains ideal for graphically intense but less interactive experiences.

Ultimately, the engine choice depends on the project's budget, goals, technical expertise, and hardware requirements. However, Unreal Engine presents several advantages in storytelling due to its high-fidelity rendering, advanced real-time lighting, and cinematic tools. Its Sequencer allows for seamless in-engine cinematics, enabling developers to craft interactive narratives that feel polished and immersive. Additionally, MetaHuman Creator provides lifelike characters with detailed facial animations, enhancing emotional engagement in storytelling-driven projects. While Unity remains an accessible and flexible choice for indie developers, Unreal Engine excels in creating narrative-rich, visually stunning experiences that push the boundaries of immersive storytelling, particularly in AAA games, virtual production, and VR applications.

\section{Challenges and Ethical Concerns}\label{CH}

Unreal Engine's advanced capabilities, while groundbreaking, come with significant challenges.

\subsection{Hardware Challenges}

Regarding hardware requirements and accessibility, key features like Nanite and Lumen demand high-performance GPUs and substantial memory resources, making the development process costly and often inaccessible to smaller teams \cite{satheesh2016unreal}. Similarly, VR content created with Unreal Engine requires users to possess high-quality hardware to fully experience its potential, raising the entry barrier and limiting access for broader audiences \cite{mack2019unreal}.
%Reviewer 2 comment 6
To address performance challenges associated with Unreal Engine, several optimization strategies have emerged, including cloud-based rendering, AI-driven performance enhancements, and hybrid computing architectures.

\subsubsection{Cloud-Based Rendering for Unreal Engine}

Cloud-based rendering offloads the computational workload to remote high-performance servers, allowing developers to create and deploy high-fidelity Unreal Engine 5 applications without requiring expensive local hardware. Several cloud solutions are emerging for Unreal Engine:

\begin{itemize}
    \item \textbf{NVIDIA GeForce Now and Amazon Luna:} These platforms provide cloud-based streaming for Unreal Engine 5 applications, enabling high-quality rendering on consumer-grade devices \cite{nvidia2025geforce, amazon2025luna}.
    
    \item \textbf{Pixel Streaming in Unreal Engine 5:} This feature allows real-time interactive experiences to be streamed from cloud GPUs to lightweight clients, such as web browsers or mobile devices \cite{epicgames2025pixel}.
    
    \item \textbf{Google Cloud GPU Instances:} Developers can leverage virtual machines with NVIDIA RTX GPUs to run Unreal Engine 5 in cloud-based workflows, reducing the strain on local hardware \cite{google2025cloud}.
\end{itemize}

Cloud-based rendering significantly reduces entry barriers for developers, enabling real-time collaboration on large projects and improving accessibility for mobile and low-power devices. However, latency, bandwidth limitations, and data security concerns must be considered when adopting cloud solutions \cite{hatami2024survey}.

\subsubsection{AI-Driven Optimization Techniques}

AI-driven rendering and performance optimization have introduced novel ways to enhance Unreal Engine 5’s efficiency:

\begin{itemize}
    \item \textbf{AI-Based LOD Generation:} Unreal Engine 5 can integrate machine-learning models to dynamically adjust the level-of-detail based on scene complexity, reducing GPU workload without sacrificing visual fidelity \cite{hassani5189246optimizing}.
    
    \item \textbf{Neural Upscaling with DLSS and FSR:} NVIDIA’s Deep Learning Super Sampling (DLSS) and AMD’s FidelityFX Super Resolution (FSR) improve frame rates by rendering at lower resolutions and using AI to upscale images in real time. These techniques provide 60--120\% performance boosts in Unreal Engine scenes \cite{nvidia2025dlss, amd2025fsr}.
    
    \item \textbf{AI-Driven Asset Optimization:} Procedural asset compression algorithms use AI to reduce texture and polygon complexity while maintaining high visual quality, allowing smoother performance on mid-range hardware \cite{koniaris2017real}.
\end{itemize}

These AI-driven techniques enable Unreal Engine 5 projects to scale dynamically based on available hardware, improving accessibility and reducing computational overhead.

\subsubsection{Hybrid Computing Architectures}

Hybrid computing combines local processing with cloud-based assistance, allowing developers to balance performance and cost. Emerging hybrid solutions for Unreal Engine include:

\begin{itemize}
    \item \textbf{Edge Computing with 5G Support:} Real-time rendering tasks can be partially offloaded to edge servers, reducing latency while keeping crucial interactions processed locally \cite{lochmann2021latency}.
    
    \item \textbf{Distributed Rendering Pipelines:} Studios can divide workloads between local GPUs and remote servers, optimizing rendering pipelines to match device capabilities \cite{li2018real}.
    
    \item \textbf{Hybrid VR Systems:} VR applications can use local tracking and physics simulations while rendering complex scenes in the cloud, making VR storytelling more accessible to users with mid-range hardware \cite{tan2024dhr+}.
\end{itemize}

These hybrid approaches bridge the gap between cloud-based convenience and local hardware control, ensuring consistent performance across multiple platforms while reducing hardware constraints.

\subsection{Accessibility Concerns}

Accessibility and inclusivity remain critical concerns in immersive storytelling. VR design that accommodates users with disabilities is still underdeveloped \cite{wilson1997virtual}. Many VR applications lack key accessibility supports such as:
\begin{itemize}
    \item \textbf{Motion Sensitivity Issues} – VR motion sickness due to high frame rates, fast camera movements, or a lack of customizable comfort settings \cite{tang2025strategies}.
    \item \textbf{Limited Input Options} – Most applications rely on standard controllers, making them inaccessible to users with mobility impairments \cite{bimba2017adaptive}.
    \item \textbf{Audio-Visual Accessibility Gaps} – Insufficient support for screen readers, closed captions, or colorblind-friendly UI designs \cite{mott2019accessible}.
\end{itemize}

\subsubsection{Recommendations for Inclusive Design in Unreal Engine 5}

To enhance accessibility, Unreal Engine 5 developers can implement the following solutions:

\begin{itemize}
   \item \textbf{Adaptive Interfaces} – Unreal Engine 5’s \textit{UMG (Unreal Motion Graphics UI Designer)} allows developers to create scalable, customizable interfaces that accommodate different user needs. Features such as resizable text, high-contrast modes, and voice navigation can improve accessibility \cite{dombrowski2019designing}.
    \item \textbf{Alternative Input Mechanisms} – Unreal Engine supports \textit{voice commands, eye-tracking, and haptic feedback}, which can be integrated to provide alternative control options for users with mobility impairments \cite{moreno2024eye}.
    \item \textbf{Motion Comfort Settings} – Developers can reduce motion sickness risks by incorporating \textit{adjustable locomotion options} (e.g., teleportation movement, vignette effects) and customizable field-of-view settings \cite{zhao2022mitigation}.
    \item \textbf{Assistive Audio Features} – Implementing \textit{spatialized audio cues, descriptive narration, and customizable subtitles} improves accessibility for users with hearing or visual impairments \cite{balan2014audio}.
    \item \textbf{XR Accessibility Standards} – Following guidelines from the \textit{XR Access Initiative} \cite{xraccess2023} and \textit{W3C Web Accessibility Initiative (WAI)} ensures compatibility with assistive technologies.
\end{itemize}

By incorporating these design principles, Unreal Engine can support more inclusive and accessible immersive experiences, allowing a broader range of users to engage with virtual environments effectively.
Options for customizing motion sensitivity or providing alternative input methods can exclude individuals with specific needs. Although Unreal Engine offers tools to create adaptive interfaces, integrating them requires additional time and effort from developers, posing further challenges \cite{lou2021hand}.

\subsection{Ethical Implications}

Ethical implications also arise from the creation of highly immersive content. Ultra-realistic environments and the evocation of instinctive emotions can lead to over-immersion, potentially distorting users' perception of time and reality \cite{bowman2007virtual} as well as influencing user behavior, emotional states, and cognitive processing \cite{madary2016real}. Moreover, VR's ability to track user behavior introduces privacy concerns, particularly regarding collecting and using personal data for targeted advertising or commercial purposes \cite{pfeuffer2019behavioural}. Addressing these challenges requires a structured approach, encompassing regulatory compliance, developer guidelines, and mitigation strategies.

\subsubsection{Regulatory Frameworks for Data Privacy and VR Ethics}

As immersive experiences collect increasing amounts of biometric and behavioral data, regulatory frameworks are crucial in ensuring ethical data practices. Regulations such as the General Data Protection Regulation (GDPR) in Europe and the California Consumer Privacy Act (CCPA) set legal precedents for protecting user information, including data gathered through eye tracking, motion sensors, and voice inputs in VR applications \cite{voigt2017eu, CCPA2025}.

Additionally, emerging policies such as the proposed EU Artificial Intelligence Act classify immersive technologies under "high-risk AI systems," suggesting stricter data governance and ethical oversight in real-time content generation and machine learning-driven storytelling \cite{EUAIAct2021}.

For Unreal Engine developers working with AI-driven storytelling, these regulations necessitate:

\begin{itemize}
    \item Transparent data handling practices.
    \item User consent mechanisms for data collection.
    \item Opt-out options for personalized content tracking.
\end{itemize}

Failure to comply with these guidelines could result in legal consequences and loss of consumer trust, particularly in VR applications involving personalized advertising, biometric tracking, or mental health simulations \cite{de2019security}.

\subsubsection{Developer Guidelines for Ethical Immersive Content Creation}

To mitigate ethical concerns, industry-standard guidelines such as those provided by the XR Safety Initiative (XRSI) and IEEE’s Ethically Aligned Design for AI \& XR offer best practices for immersive storytelling \cite{XRSI, IEEEAutonomous2025}. These include:

\begin{itemize}
    \item \textbf{Transparency in Virtual Environments}: Informing users when interacting with AI-generated NPCs, synthetic media, or algorithm-driven narratives \cite{cave2020ai}.
    \item \textbf{Content Moderation for Psychological Safety}: Implementing ethics-driven procedural storytelling to avoid distressing over-immersion, particularly in VR simulations replicating trauma or violence \cite{madary2016real}.
    \item \textbf{Fair Representation and Accessibility}: Ensuring AI-generated characters and narratives do not perpetuate stereotypes, bias, or exclusionary practices in digital storytelling \cite{raji2020closing}.
\end{itemize}

% Review 2 Reviewer 2 comment 4
In addition to realism-related concerns, emerging ethical questions focus on user data sensitivity and the implications of AI-generated content in narrative design. As immersive environments increasingly rely on biometric and behavioral data, such as gaze tracking, voice input, and physical movement, there is a heightened need for transparency in how this information is processed and used in real-time adaptive storytelling\cite{slater2020ethics}.

Moreover, narratives dynamically constructed by AI systems may obscure their synthetic origins, potentially misleading users about the source of emotional content or character agency\cite{cave2023imagining}. This raises important questions regarding authorship, informed consent, and the ethical obligation to disclose when content is algorithmically generated. Developers are encouraged to ensure adaptive narratives remain interpretable and respectful of user autonomy, reinforcing trust without diminishing creative flexibility\cite{mittelstadt2016ethics}.

%--
Unreal Engine’s ability to create lifelike, emotionally expressive AI characters introduces ethical dilemmas regarding realism and deception in media. Developers must consider whether AI-driven NPCs should disclose their non-human nature and how emotionally manipulative interactions could impact user well-being \cite{zhuk2024ethical}.

\subsubsection{Mitigation Strategies for Over-Immersion and Privacy Risks}

While immersive experiences enhance engagement, excessive immersion may lead to dissociation from reality or neurological fatigue \cite{slater2016enhancing}. Unreal Engine developers must consider content design choices that prevent negative psychological effects:

\textbf{Preventing Over-Immersion:}
\begin{itemize}
    \item \textbf{Session Time Limits and Break Reminders}: Integrating adaptive break systems that detect prolonged VR sessions and prompt users to rest, reducing mental fatigue \cite{riva2019neuroscience}.
    \item \textbf{Cognitive Load Management}: Implementing gradual difficulty curves and adaptive learning mechanisms to ensure users do not experience sensory overload in high-intensity storytelling environments \cite{bowman2007virtual}.
    \item \textbf{Ethical Use of Behavioral Analytics}: Developers should avoid dark patterns in game mechanics that manipulate user engagement (e.g., endless loops or reward-based compulsion systems), which can contribute to addictive behaviors and reduced autonomy \cite{zagal2013dark}.
\end{itemize}

\textbf{Enhancing Privacy Controls:}
To mitigate privacy risks, Unreal Engine’s data security framework should include:
\begin{itemize}
    \item \textbf{On-device AI processing instead of cloud-based tracking} to reduce exposure to third-party data collection \cite{vosoughi2018spread}.
    \item \textbf{Encrypted user sessions} for VR applications that involve sensitive data (e.g., mental health simulations, biometric interactions) \cite{ramolia2025futuristic}.
\end{itemize}

Addressing these concerns ensures that Unreal Engine remains a trusted platform for developers, users, and regulatory bodies, fostering responsible innovation in immersive storytelling and VR applications.

\section{Future Directions}\label{F}

The future of Unreal Engine in immersive storytelling and virtual reality is poised to evolve alongside rapid technological advancements and the ever-changing ideas of creators. A pivotal area in this evolution is the integration of AI systems. While some games already feature AI-driven NPCs, the implementation of AI in gaming remains limited in scope \cite{boyd2017reinforcement}. Future applications could include dynamic, procedurally generated content and adaptive storytelling. For instance, AI could enable developers to create storylines that adapt to user behavior throughout the game, resulting in deeply personalized experiences.
%Reviewer 2 comment 10
An exemplary case of this evolution is the upcoming game \textit{Marvel Rivals}, developed using Unreal Engine 5. As a highly anticipated AAA title, \textit{Marvel Rivals} showcases the potential of real-time physics simulations and destructible environments. Utilizing the Chaos Physics system, the game delivers unprecedented environmental destruction, allowing players to dynamically interact with and alter the world. Buildings collapse realistically, debris scatters in response to force, and objects respond naturally to player actions. This level of interactivity significantly enhances immersion and opens new possibilities for gameplay mechanics \cite{marvel2025rivals}. 

\begin{figure}[h]
    \centering
    \includegraphics[width=0.45\textwidth]{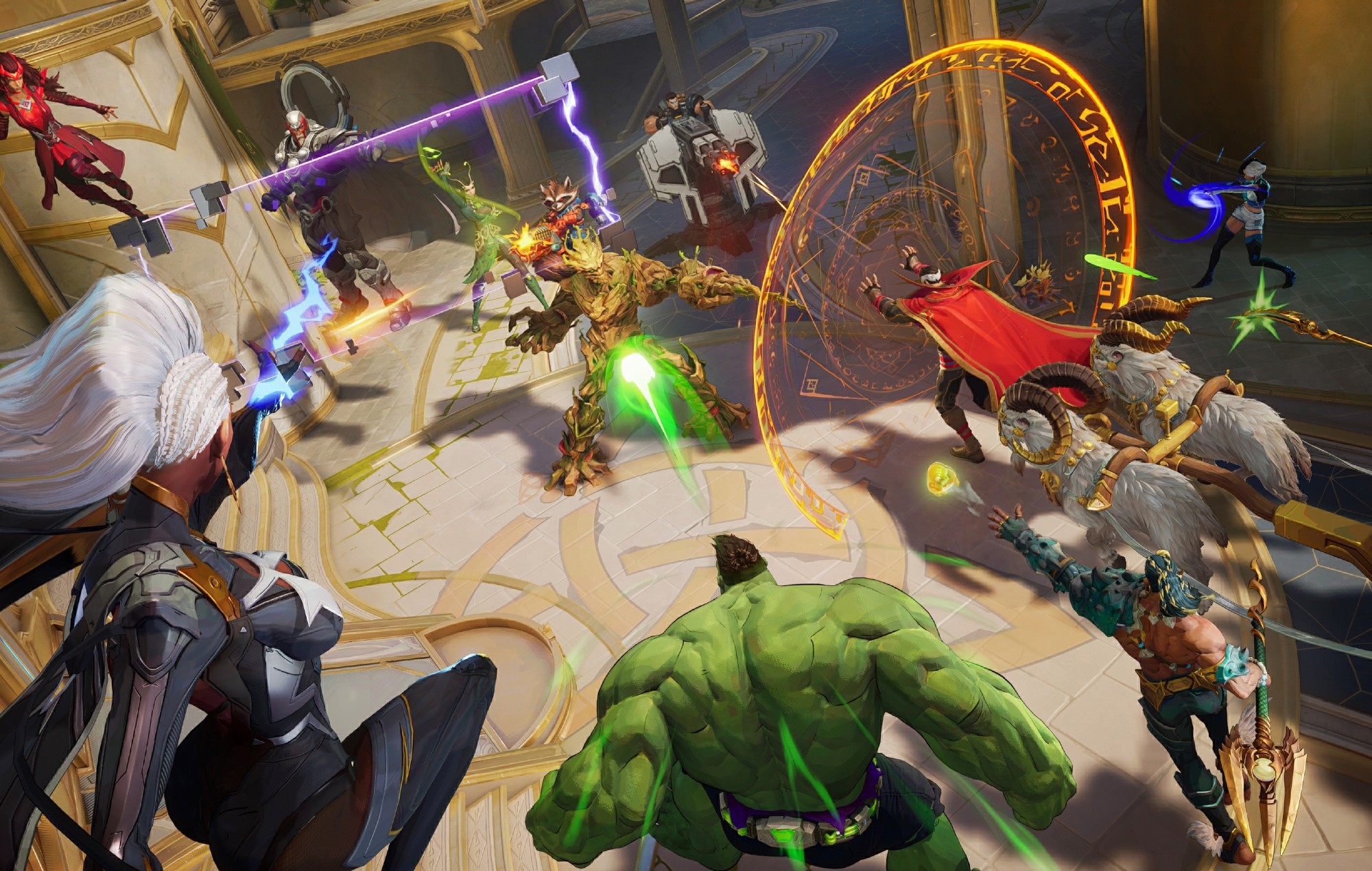}
    \caption{Screenshot from \textit{Marvel Rivals}, showcasing environments rendered with Unreal Engine 5's Chaos Physics and Nanite \cite{marvel2025rivals}.}
    \label{fig:marvel_rivals}
\end{figure}

Furthermore, the integration of Nanite ensures that the visual quality remains high even in scenes with numerous fragmented objects and complex geometry. Real-time lighting via Lumen adds to the realism by dynamically adapting to environmental changes, such as light streaming through destroyed structures. As illustrated in Figure~\ref{fig:marvel_rivals}, the game leverages Unreal Engine 5's real-time rendering to maintain high visual fidelity, while the destructible elements and interactive surroundings create an immersive and chaotic battlefield.

These innovations demonstrate the advanced technical capabilities of Unreal Engine 5 but also highlight its potential to enhance gameplay through dynamic, interactive worlds. By leveraging advanced physics simulation and real-time rendering, games like \textit{Marvel Rivals} set a new standard for realism and immersion in interactive media.

\subsection{AI-Driven Workflows and Procedural Content Generation}
% Review 2 - Reviewer 2 comment 8
AI is transforming content creation in Unreal Engine, particularly in procedural content generation, adaptive storytelling, and automated asset creation. By leveraging AI-driven tools, developers can streamline workflows and enhance user immersion through dynamic, evolving environments. Tools like Houdini Engine, PCG Graph, and World Partition enable context-aware generation of modular terrain, architecture, and asset layouts \cite{beal2025fabrination}. These systems reduce manual workload and support scalable world design for open-world games, simulations, and training applications.

\textbf{Procedural content generation (PCG)} utilizes AI algorithms to create vast, complex environments with minimal manual effort. Techniques such as \textit{rule-based generation}, \textit{neural networks}, and \textit{genetic algorithms} allow Unreal Engine to automatically generate landscapes, architecture, and level layouts \cite{togelius2011search}. For example, AI-powered terrain synthesis tools can produce detailed open-world environments by analyzing geospatial data and optimizing memory usage \cite{smelik2009survey}. Recent implementations include Unreal Engine’s PCG Graph and integration with Houdini Engine, which allow designers to automate modular asset placement, terrain variation, and environmental storytelling based on spatial and narrative constraints \cite{westerlund2022evaluating}.

\textbf{Adaptive storytelling} enables narratives to evolve based on player behavior. Machine learning models analyze user choices and dynamically adjust dialogues, character interactions, and quest structures in real-time \cite{riedl2016computational}. This creates \textit{personalized narrative experiences}, improving engagement and replayability. Unreal Engine's Blueprints system, integrated with AI-driven logic, allows for complex decision trees that change based on user interactions, fostering non-linear storytelling. Tools such as Inworld AI and Convai, which leverage transformer-based models, create responsive NPCs capable of maintaining coherent, player-aware dialogue \cite{ziegler2019fine, freiknecht2020procedural}. These systems are increasingly applied in immersive experiences, including open-world games and metaverse platforms like Unreal Editor for Fortnite (UEFN) \cite{epicgames2023uefn}.

\textbf{Automated asset creation} further enhances development efficiency. AI-powered texture synthesis, animation retargeting, and character generation streamline production workflows. Tools like \textit{MetaHuman Creator} allow developers to generate lifelike NPCs with minimal effort, reducing the time required for detailed character modeling \cite{thies2020neural}. Similarly, AI-assisted motion capture refines character animations, making them more realistic and reducing the need for extensive manual adjustments \cite{dathathri2019plug}.

%--
\subsection{Recommendations}

To maintain its leadership in immersive storytelling, Unreal Engine researchers and developers should focus on the following priorities:

\begin{itemize}
    \item \textbf{Hardware Accessibility:} Investigate and implement cloud-based rendering solutions to minimize hardware requirements for developers and users, reducing barriers to entry \cite{nowak2018modeling}.
    
    \item \textbf{Inclusive Design:} Enhance accessibility by developing customizable interfaces and input methods as standard features in VR applications. These should cater to users with varying levels of technical proficiency and those with disabilities or mobility challenges.
    
    \item \textbf{Ethical Considerations:} Establish transparent data privacy practices for VR environments that collect user interaction data. Additionally, address over-immersion risks by incorporating thoughtful content design and clearly defined boundaries.
    
    \item \textbf{Interdisciplinary Collaboration:} Encourage partnerships between technologists, artists, writers, and researchers to push the boundaries of interactive storytelling and VR design, fostering innovation across disciplines.
    
    \item \textbf{AI Integration:} Further develop AI-driven tools for procedural content generation, dynamic storytelling, and automated asset creation to enhance immersive experiences.
\end{itemize}

By addressing these priorities, Unreal Engine can continue to lead the way in immersive storytelling, ensuring its technology remains inclusive, ethical, and accessible while pushing the limits of creativity and innovation.

\section{Conclusion}\label{C}

Unreal Engine is a versatile development environment that has redefined the landscape of immersive storytelling and virtual reality. By integrating advanced technologies such as Nanite for high-detail asset rendering and Lumen for dynamic global illumination, the engine has empowered creators to produce visually stunning and emotionally engaging experiences across various applications. From gaming and virtual production to architectural visualization and cultural preservation, Unreal Engine's adaptability and scalability have enabled innovation and creativity in ways previously unattainable.

This paper highlights how tools like the MetaHuman Creator, Sequencer, and Blueprints simplify complex workflows while maintaining high levels of fidelity and interactivity. These features have made Unreal Engine a cornerstone for interdisciplinary collaboration, merging technological precision with artistic expression to craft narratives that resonate deeply with users.

However, the engine's high hardware requirements present a significant barrier for smaller teams and broader accessibility, particularly in VR applications. Unreal Engine's advanced capabilities also bring forth critical concerns regarding inclusivity, accessibility for users with disabilities, and the ethical implications of over-immersion and data privacy. Addressing these challenges is paramount to ensuring the platform remains equitable and sustainable for all users.

The future of Unreal Engine lies in harnessing emerging technologies such as AI and cloud-based rendering to expand its reach and accessibility. By integrating dynamic, procedurally generated content and adaptive storytelling mechanisms, Unreal Engine can deliver even more personalized and impactful user experiences. Furthermore, prioritizing ethical design and fostering interdisciplinary partnerships will be essential for pushing the boundaries of immersive storytelling while maintaining responsible innovation.

Unreal Engine is more than a technical tool; it supports creative experimentation and adaptive storytelling. Its contributions to immersive media have set a new benchmark in visual fidelity, interactivity, and emotional engagement. With continuous advancements and a commitment to addressing existing challenges, Unreal Engine is poised to remain a pivotal force in shaping the future of virtual reality and interactive media.

\section*{Acknowledgment}
This research was financially supported by the TUM Campus Heilbronn Incentive Fund 2024 of the Technical University of Munich, TUM Campus Heilbronn. We gratefully acknowledge their support, which provided the essential resources and opportunities to conduct this study.

\balance
\bibliographystyle{elsarticle-num} 
\bibliography{Storytelling}

\end{document}